\documentclass[aps,prb,twocolumn]{revtex4}
\usepackage{graphics}

\begin{document}

\title{SO(5) superconductor in a Zeeman magnetic field:\\phase diagram and thermodynamic properties}

\author{T. A. Zaleski}
\author{T. K. Kope\'{c}}
\affiliation{Institute of Low Temperature and Structure Research,\\Polish Academy of %
Sciences,\\P.O. Box 1410, 50-950 Wroc\l aw, Poland}

\begin{abstract}
In this paper we present calculations of the SO(5)\ quantum rotor theory of
high-T$_{c}$ superconductivity in Zeeman magnetic field. We use the
spherical approach for five-component quantum rotors in three-dimensional
lattice to obtain formulas for critical lines, free
energy, entropy and specific heat and present temperature dependences of
these quantities for different values of magnetic field. Our results are in
qualitative agreement with relevant experiments on high-T$_{c}$ cuprates.
\end{abstract}


\maketitle

\section{Introduction}

The SO(5)\ theory of high-T$_{c}$ superconductivity, unifying antiferromagnetism 
and superconductivity was proposed to combine SO(3) antiferromagnetic (AF) 
staggered magnetization with superconducting (SC) two-component U(1) real paring 
field into a new order parameter.\cite{ZhangScience} The smallest symmetry group 
meeting this requirements is the SO(5) group. It contains SO(3) group of spin 
rotation and U(1) gauge group of \textquotedblleft charge rotation\textquotedblright %
as subgroups along with additional, the so-called \textquotedblleft 
$\pi $\textquotedblright  operators rotating AF to SC state and vice versa. In the 
SO(5)\ theory ordered phases arise once SO(5)\ symmetry is spontaneously broken: 
with SO(3) symmetry breaking AF phase arises, while for U(1) symmetry breaking SC 
phase appears. Consequently, AF\ and SC order parameters are grouped in five component
vector called \textquotedblleft superspin\textquotedblright , which direction is 
related to the competition between AF\ and SC states and the kinetic energy of
the system is that of a SO(5) quantum rigid rotor. The low-energy dynamics is 
determined in terms of the Goldstone modes and their interactions specified 
by the SO(5) symmetry. 

Although, the SO(5) theory was originally proposed in the context of an effective 
quantum non-linear $\sigma $ model (NLQ$\sigma $M) description of the SO(5) rotor 
model its predictions were tested also within microscopic models. \cite
{MicroscopicModels1,MicroscopicModels2,MicroscopicModels3,MicroscopicModels4,%
MicroscopicModels5,Arrigoni} Calculations based on NLQ$\sigma $M showed that the 
features of the phase diagram deduced from the SO(5)\ theory agree qualitatively 
with the global topology of the observed phase diagrams of high-T$_{c}$ superconductors. 
\cite{PRB1} Magnetic correlation functions within the SO(5)\ theory were 
investigated showing that the theory predicts a scenario for the evolution of
magnetic behaviour, which is qualitatively consistent with experiments. \cite{PRB2} 
Furthermore, the study of the quantum critical point scenario within
the concept of the SO(5) group showed that the scaling of the
contribution to the electrical resistivity due to spin fluctuations, displayed
a linear resistivity dependence on temperature for increasing quantum
fluctuations, which is a hallmark example of anomalous properties of cuprate
materials.\cite{PRL} Finally, thermodynamic properties of the SO(5) model were
studied, where entropy and specific heat 
were calculated and compared with experimental findings.\cite{JPHYS}

Many experimentally observed properties of high-T$_{c}$ cuprates show strong
dependence on the magnetic field, e.g. entropy, specific heat, magnetic susceptibility,
electrical resistivity, etc. Therefore, a proper theory of
high-T$_{c}$ superconductors must be able to explain the magnetic properties of these 
materials. Since, SO(5) model in the presence
of finite chemical potential and a finite Zeeman magnetic field has the 
exact SU(2)$\times $U(1) symmetry, the Zeeman magnetic field can be
introduced, and the SO(5) theory can be tested in any doping level.\cite
{HuZhang}

The aim of this paper is to study influence of the Zeeman
magnetic field within the SO(5)\ theory. Obtained results (e.g.
specific heat) can be then (qualitatively) compared with the relevant experiments and test 
the basic principles of the SO(5) theory.

The outline of the reminder of the paper is as follows. In Section II we
introduce the quantum SO(5)\ Hamiltonian in the Zeeman magnetic field and find 
the corresponding Lagrangian of NLQ$\sigma $M. In Section III we establish the 
phase diagram of the system in an applied magnetic field. Section IV is devoted 
to the study of magnetic dependence of various thermodynamic functions: free 
energy, entropy and the specific heat. Finally, in Section V we summarize the conclusions to be 
drawn from our work.

\section{Hamiltonian and the effective Lagrangian}

We start from the low-energy Hamiltonian of superspins $\mathbf{n}_{i}$
placed in the discrete 3D s.c. lattice
(3DSC) in the Zeeman magnetic field $B$ along $y$ axis. The sites are numbered
by indices $i$ and $j$ running from $1$ to $N$ -- the total number of sites.
The superspin components, labeled by $\mu $ and $\nu =1,...,5$ refer to AF ($%
\mathbf{n}_{i}^{AF}=\left( n_{2},n_{3},n_{4}\right) _{i}$) and SC ($\mathbf{n%
}_{i}^{SC}=\left( n_{1},n_{5}\right) _{i}$) order parameters, respectively. The
superspin components are mutually commuting variables\cite{ZhangScience} and their values are 
restricted by the rigid rotator constraint $\mathbf{n}_{i}^{2}=1$. The SO(5) Hamiltonian 
\begin{eqnarray}
H &=&\frac{1}{2u}\sum_{i}\sum_{\mu <\nu }L_{i}^{\mu \nu }L_{i}^{\mu \nu
}-\sum_{i<j}J_{ij}\mathbf{n}_{i}\cdot \mathbf{n}_{j}+  \nonumber \\
&&-V\left( \mathbf{n}_{i}\right) -B\sum_{i}L_{i}^{24}-2\mu \sum_{i}L_{i}^{15}
\label{Eq_Hamiltonian}
\end{eqnarray}
consists of three parts:\ the kinetic energy of the rotors (where $u$ is an
analogue of moment of inertia), intersite interaction energy (with $J_{ij}$ being 
the stiffness in the charge and spin channel) and SO(5) symmetry breaking part 
(including the Zeeman magnetic field acting in AF sector). 
The quantities 
$
L_{i}^{\mu \nu }=n_{\mu i}p_{\nu i}-n_{\nu i}p_{\mu i} 
$
are generators of the Lie SO(5) algebra (related to the total charge, spin
and so-called `$\pi $' operators\cite{ZhangScience}) and $p_{\mu i}$ are linear 
momenta given by: 
\begin{eqnarray}
&&p_{\mu i}=i\frac{\partial }{\partial n_{\mu i}},  \nonumber \\
&&\left[ n_{\mu },p_{\nu }\right] =i\delta _{\mu \nu }\text{.}
\end{eqnarray}
Furthermore, we assume that $J_{ij}\equiv J(\mathbf{R}_{i}-\mathbf{R}_{j})$ is 
nonvanishing for the nearest neighbours and its Fourier transform 
\begin{equation}
J_{\mathbf{q}}=\frac{1}{N}\sum_{\mathbf{R}_{i}}J\left( \mathbf{R}_{i}\right)
e^{-i\mathbf{R}_{i}\cdot \mathbf{q}}
\end{equation}
is simply $J_{\mathbf{q}}=J\epsilon (\mathbf{k})$, where
\begin{equation}
\epsilon_{\mathbf{q}}=\cos q_{x}+\cos q_{y}+\cos q_{z}
\label{Eq_Jq}
\end{equation}
is the structure factor for the 3D s.c. lattice.\cite{3d} 

The last three parts of the Hamiltonian provide SO(5)\ symmetry breaking
terms. In the result of their interplay, the system favours either the `easy
plane' in the SC space $\left( n_{1},n_{5}\right) $, or an `easy sphere' in
the AF space $\left( n_{2},n_{3},n_{4}\right) $. Two of the three terms
influence directly AF order parameter: 
\begin{equation}
V\left( \mathbf{n}_{i}\right) -B\sum_{i}L_{i}^{24}=\frac{w}{2}\sum_{i}\left(
n_{2i}^{2}+n_{3i}^{2}+n_{4i}^{2}\right) -B\sum_{i}L_{i}^{24},
\end{equation}
where $w$ is the anisotropy constant, $B$ is the Zeeman magnetic field and $%
L_{i}^{24}$ is $y$ component of the spin vector. Positive values of $w$ and $%
B$ favour the AF state. The remaining term acts on SC sector and contains
total charge operator $L_{i}^{15}$, whose expectation value yields the
doping concentration and the chemical potential $\mu $ (measured from
half-filling), which positive value fovours the SC state.

We express the partition function $Z=Tre^{-\beta H}$ using the functional
integral in the Matsubara \textquotedblleft imaginary time\textquotedblright  $\tau $ formulation \cite{PRB1}
($0\leq \tau \leq 1/k_{B}T\equiv \beta $, with $T$ being the temperature). Explicitly:
\begin{eqnarray}
Z &=&\int \prod_{i}\left[ D\mathbf{n}_{i}\right] \int \prod_{i}\left[ \frac{D%
\mathbf{p}_{i}}{2\pi }\right] \delta \left( 1-\mathbf{n}_{i}^{2}\right)
\delta \left( \mathbf{n}_{i}\cdot \mathbf{p}_{i}\right) \times  \nonumber \\
&\times &\exp \left\{ -\int_{0}^{\beta }d\tau \left[ i\mathbf{p}\left( \tau
\right) \cdot \frac{d}{d\tau }\mathbf{n}\left( \tau \right) +H\left( \mathbf{%
n},\mathbf{p}\right) \right] \right\} =  \nonumber \\
&=&\int \prod_{i}\left[ D\mathbf{n}_{i}\right] \delta \left( 1-\mathbf{n}%
_{i}^{2}\right) e^{-\int_{0}^{\beta }d\tau \mathcal{L}\left( \mathbf{n}%
\right) }  \label{Eq_PartitionFunction}
\end{eqnarray}
with $\mathcal{L}$ being the Lagrangian:
\begin{eqnarray}
\mathcal{L}\left( \mathbf{n}\right) &=&\frac{1}{2}\left[ \sum_{i}u\left( 
\frac{\partial \mathbf{n}_{SC}}{\partial \tau }\right) ^{2}+u\left( \frac{%
\partial \mathbf{n}_{AF}}{\partial \tau }\right) ^{2} \right. + \nonumber \\
&-&4u\mu ^{2}\mathbf{n}_{SC}^{2}+4iu\mu \left( \frac{\partial n_{1}}{\partial \tau }n_{5}-\frac{%
\partial n_{5}}{\partial \tau }n_{1}\right) +  \nonumber \\
&-&uB^{2}\left( n_{2}^{2}+n_{4}^{2}\right) +2iuB\left( \frac{\partial
n_{2}}{\partial \tau }n_{4}-\frac{\partial n_{4}}{\partial \tau }%
n_{2}\right) + \nonumber \\
&-&\left. \sum J_{ij}\mathbf{n}_{i}\cdot \mathbf{n}_{j}-w\sum_{i}\left(
n_{2i}^{2}+n_{3i}^{2}+n_{4i}^{2}\right) \right] \text{.}
\label{Eq_Lagrangian}
\end{eqnarray}
The problem can be solved exactly in terms of the spherical model.\cite{Sphere} 
We note that the superspin rigidity constraint ($\mathbf{n}_{i}^{2}=1$) implies that a
weaker condition also holds, namely: 
\begin{equation}
\frac{1}{N}\sum_{i=1}^{N}\mathbf{n}_{i}^{2}=1\text{.}  \label{SphericalConstraint}
\end{equation}

Therefore, the superspin components $\mathbf{n}_{i}\left( \tau \right) $ satisfying the 
quantum periodic boundary condition $\mathbf{n}_{i}\left( \beta \right) =\mathbf{n}_{i}\left( 0\right)$
will be treated as \textit{continuous} variables, i.e., $-\infty <\mathbf{n}%
_{i}\left( \tau \right) <\infty $, but constrained on average (due to Eq. (%
\ref{SphericalConstraint})). The constraint can be implemented using Dirac-$\delta $ function 
$\delta \left( x \right) = \int_{ - \infty }^{ + \infty } {[ {{d\lambda }}/{{2\pi }}] e^{i\lambda x} }$
, which introduces the Lagrange multiplier $\lambda \left( \tau \right) $
adding an additional quadratic term (in $\mathbf{n}_{i}$ fields) to the
Lagrangian (\ref{Eq_Lagrangian}). Consequently, the partition function
reads: 
\begin{equation}
Z=\int \left[ \frac{d\lambda }{2\pi i}\right] e^{-N\phi \left( \lambda \right) }\text{,}
\end{equation}
where the function $\phi \left( \lambda \right) $ is defined as: 
\begin{eqnarray}
\phi \left( \lambda \right) &=&-\int_{0}^{\beta }d\tau \lambda \left( \tau
\right) -\frac{1}{N}\ln \int \prod_{i}\left[ D\mathbf{n}_{i}\right] \times 
\nonumber \\
&&\times \exp \left[ -\sum_{i}\int_{0}^{\beta }d\tau \left( \mathbf{n}%
_{i}^{2}\lambda \left( \tau \right) -\mathcal{L}\left[ \mathbf{n}\right]
\right) \right] \text{.}  \label{PhiFunction}
\end{eqnarray}
In the thermodynamic limit ($N\rightarrow \infty $), the method of steepest descent is exact and
the saddle point $\lambda \left( \tau \right) =\lambda _{0}$ satisfies the
condition: 
\begin{equation}
\left. \frac{\delta \phi \left( \lambda \right) }{\delta \lambda \left( \tau
\right) }\right| _{\lambda =\lambda _{0}}=0\text{.}  \label{SaddlePoint}
\end{equation}
At the antiferromagnetic and superconducting phase transition boundaries the corresponding order susceptibilities
become infinite (see, Ref. \onlinecite{PRB1}), which implies for the Lagrange multipliers: 
\begin{eqnarray}
\lambda _{0}^{AF} &=&\frac{1}{2}J_{\mathbf{k}=0}+\frac{w}{2}+\frac{uB^{2}}{2}, 
\nonumber \\
\lambda _{0}^{SC} &=&\frac{1}{2}J_{\mathbf{k}=0}+2u\mu ^{2},
\label{LambdaZero}
\end{eqnarray}
for AF\ and SC critical lines, respectively.

\section{Phase diagram}

Providing the spherical condition (\ref{SphericalConstraint}) with values of
the Lagrange multipliers (\ref{LambdaZero}) one can finally arrive at the
expression for the critical lines separating AF (or SC) and QD (quantum
disordered) states: 
\begin{eqnarray}
1&=&\frac{1}{2u}\int_{-\infty }^{\infty }\rho \left( \xi \right) d\xi
\left\{ \frac{\cosh \left[ \frac{\beta }{2}A\left( \xi \right) \right] }{A\left( \xi
\right) }+\right. \nonumber \\
&+&\frac{\cosh \left[ \frac{\beta }{2}D_{-}\left( \xi \right) \right] }{A\left( \xi
\right) }|\frac{\cosh \left[ \frac{\beta }{2}D_{+}\left( \xi \right) \right] }{A\left( \xi
\right) }+  \nonumber \\
&+&\left. \frac{\cosh \left[ \frac{\beta }{2}B_{-}\left( \xi \right) \right] }{C\left( \xi
\right) }+\frac{\cosh \left[ \frac{\beta }{2}B_{+}\left( \xi \right) \right] }{C\left( \xi
\right) }\right\} ,
\end{eqnarray}
where: 
\begin{equation}
\begin{tabular}{p{1.6in}p{1.6in}}  
$A\left( \xi \right) =\sqrt{\frac{2\lambda _{0}-J\xi -w}{u}},$ & $C\left(
\xi \right) =\sqrt{\frac{2\lambda _{0}-J\xi }{u}},$ \\ 
$B_{-}\left( \xi \right) =\sqrt{\frac{2\lambda _{0}-J\xi }{u}}-2\mu ,$ & $%
D_{-}\left( \xi \right) =\sqrt{\frac{2\lambda _{0}-J\xi -w}{u}}-B,$ \\ 
$B_{+}\left( \xi \right) =\sqrt{\frac{2\lambda _{0}-J\xi }{u}}+2\mu ,$ & $%
D_{+}\left( \xi \right) =\sqrt{\frac{2\lambda _{0}-J\xi -w}{u}}+B,$
\end{tabular}
\label{ABBDefs}
\end{equation}
and $\lambda _{0}=\lambda _{0}^{AF}$ ($\lambda _{0}^{SC}$) for AF (SC) line. For convenience, 
in order to perform momentum integration over the 3D Brillouin zone, we have introduced the 
density of states
\begin{equation}
\rho \left( \xi \right) =\frac{1}{N}\sum_{\mathbf{q}}\delta \left[ \xi -\epsilon(\mathbf{q})\right] \text{,}  \label{Eq_DOS}
\end{equation}
(for explicit formula, see Ref. [\onlinecite{PRB1}]). 

Regions of AF\ and SC phases are separated by the first order transition line
(for $\mu =\mu _{c}$) given by the condition of equality of free energies of
both states, which reads: 
\begin{equation}
\lambda _{0}^{SC}=\lambda _{0}^{AF}\Rightarrow \mu _{c}^{2}=\frac{w}{4u}+%
\frac{B^{2}}{4}.  \label{MuC}
\end{equation}

It is instructive to present the phase diagram as a function of physically 
measured quantity--the charge concentration $x$ instead of the chemical potential. 
In the SO(5)\ theory, the charge concentration can be deduced from the free energy: 
\begin{equation}
x=\left\langle L^{15}\right\rangle =-\frac{1}{2}\frac{df}{d\mu }=-\frac{1}{%
2\beta }\frac{d\phi \left( \lambda _{0}\right) }{d\mu }. 
\end{equation}

The temperature-charge concentration phase diagram is depicted in Fig. 1.
Here, AF to SC transition line splits into a region of constant chemical
potential, where AF and SC states coexist (mixed region M). This behaviour can
be explained by a two-phase mixture with different densities at first-order
phase transition. In this case the system globally phase-separates in two
different spatial regions with different charge densities, but the same free
energy. As a result, the added charges only change the proportion of the
mixture of the two phases, bun not the free energy, which implies infinite 
compressibility (defined as $dx/d\mu $). 

\section{Thermodynamic functions}

\subsection{Free energy}

The free energy is defined as $f=-\left( \beta N\right) ^{-1}\ln Z=\left(
\beta \right) ^{-1}\phi \left( \lambda _{0}\right) $. Using the formula (\ref
{PhiFunction}), it can be explicitly written: 
\begin{eqnarray}
f= -\lambda +F_{1}\left( A\right) &+& F_{1}\left( B_{-}\right)+
F_{1}\left( B_{+}\right)+ \nonumber\\
&+&F_{1}\left( D_{-}\right)+F_{1}\left( D_{+}\right) , \label{FreeEnergy_Final}
\end{eqnarray}
where the function:
\begin{equation}
F_{1}\left(\Xi \right) = \frac{1}{\beta }\int_{-\infty }^{\infty }\rho \left( \xi
\right) d\xi \ln \left\{ 2\sinh \left[ \frac{\beta }{2}\Xi \left( \xi \right) \right] \right\} 
\end{equation}
and $A\left( \xi \right) $, $B_{-}\left( \xi \right) $, $B_{+}\left( \xi
\right) $, $D_{-}\left( \xi \right) $ and $D_{+}\left( \xi \right) $ 
are defined by formulas (\ref{ABBDefs}).

\subsection{Entropy}

The entropy is defined as $S=k_{B}\beta ^{2}\frac{\partial f}{\partial \beta 
}$. Using the formula (\ref{FreeEnergy_Final}) we obtain:

\begin{eqnarray}
S\left( \beta \right) = F_{2}\left( A\right)&+&F_{2}\left( B_{-}\right)+
F_{2}\left( B_{+}\right)+ \nonumber\\
&+&F_{2}\left( D_{-}\right)+F_{2}\left( D_{+}\right),
\end{eqnarray}
where the function:
\begin{eqnarray}
F_{2}\left( \Xi \right) &=&\frac{k_{B}}{2}\int_{-\infty }^{\infty }\rho \left(
\xi \right) d\xi  \left( \beta \Xi \left( \xi \right) \coth \left[ \frac{\beta 
}{2}\Xi \left( \xi \right) \right] \right. + \nonumber \\
&-&\left. 2\ln \left\{ 2\sinh \left[ \frac{\beta }{2}\Xi \left( \xi \right) \right] \right\} \right) .
\end{eqnarray}

\subsection{Specific heat}

The specific heat at constant volume is defined: 
\begin{eqnarray}
C &=&-k_{B}\beta ^{2}\frac{\partial ^{2}}{\partial \beta ^{2}}\left( \beta
f\right) =-k_{B}\beta ^{2}\left\{ 2\frac{\partial f}{\partial \beta }+\beta 
\frac{\partial ^{2}f}{\partial \beta ^{2}}\right.  \nonumber \\
&&\left. +\beta \frac{d\lambda }{d\beta }\left[ \frac{\partial ^{2}f}{%
\partial \lambda ^{2}}\frac{d\lambda }{d\beta }+2\frac{\partial ^{2}f}{%
\partial \lambda \partial \beta }\right] \right\} .
\end{eqnarray}
The temperature dependence of the specific heat in various magnetic field is
depicted in Fig. 2. For low temperatures $C\left( T\right) $ is independent
on $B$ and can be approximated by $C\left( T\right) \sim T^{3}$. For higher
temperatures the dependence become roughly linear with higher values of $%
C\left( T\right) $ for higher magnetic fields. The critical temperature is depressed 
for higher fields, which is in agreement with experimental findings.
Only, the size of finite jump occurring at the critical point, seems to be
independent on $B$ (for Y-123 jump of specific heat in critical temperature
is absent because of flux lattice melting, see Ref. \onlinecite{Junod1,
Junod2}). Consequently, Zeeman magnetic field does not influence the value
of the critical exponent $\alpha =0$. In high temperatures $C\left( T\right) 
$ dependence saturates.

\section{Final remarks}

In conclusion, we have considered the influence of the Zeeman magnetic field in the SO(5)\
theory of high-T$_{c}$ superconductivity proposed by Zhang. Experimentally, in high-T$_{c}$ cuprates
the Zeeman magnetic field can be realized by applying a magnetic field parallel to
\textit{ab} planes. Using non-linear
quantum $\sigma $ model and the spherical approximation we have found
explicit expressions for the critical lines and various thermodynamic
functions (free energy, entropy and specific heat). Obtained results are in
qualitative agreement with experimentally observed features of high-T$_{c}$
cuprates. The maximum of the specific heat is depressed and shifted towards lower 
temperature in higher magnetic fields. However, it should be stressed 
that the case of the Zeeman field considered in 
the present paper neglects the orbital effects, which are important in high-T$_{c}$ cuprates,
when the magnetic field is applied perpendicularly to \textit{ab} planes.

\section{Acknowledgment}

This work was supported by Polish State Committee for Scientific Research
grant No. 2PO3B04922.


\begin{figure}[tbh]
\scalebox{0.6}{\includegraphics{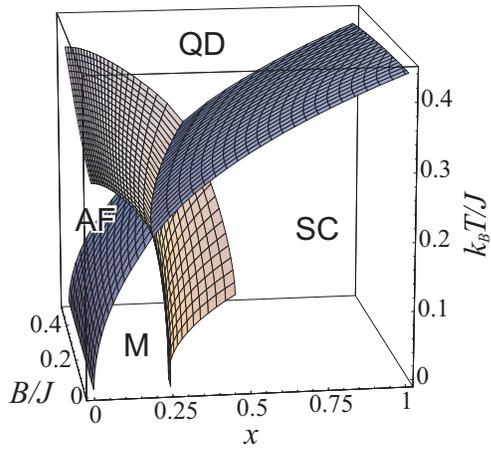}}
\caption{$T-x-B$ phase diagram for $uJ=3$ and $w/J=1$.}
\label{Fig_PDiagram2}
\end{figure}

\begin{figure}[tbh]
\scalebox{0.6}{\includegraphics{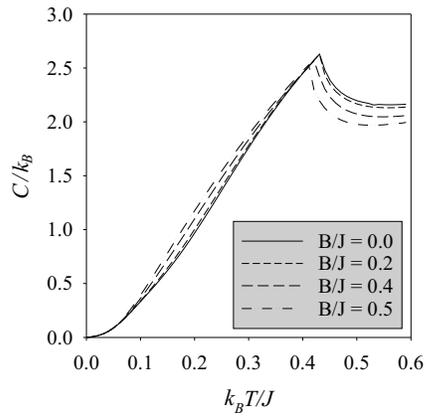}}
\caption{Temperature dependence of specific heat for various magnetic
fields, for $uJ=3$ and $w/J=1$.}
\label{Fig_SpecHeat}
\end{figure}

\end{document}